\documentclass[aps,pra,showpacs,twocolumn,final,floatfix]{revtex4}

\usepackage{amsmath,float,graphicx,subfigure,array}
\begin{document}

\title{Enhancing quantum coherence with short-range correlated disorder}
\author{P. Capuzzi}
\email{capuzzi@df.uba.ar}

\affiliation{Departamento de F{\'\i}sica, Universidad de Buenos Aires,
  Ciudad Universitaria, Buenos Aires 1428, Argentina}

\affiliation{Instituto de F{\'\i}sica de Buenos Aires, CONICET, Ciudad
  Universitaria, Buenos Aires 1428, Argentina}

\author{M. Gattobigio}
\email{Mario.Gattobigio@inln.cnrs.fr}
\affiliation{Institut Non Lin\'eaire de Nice UMR7335,
Universit\'e de Nice-Sophia Antipolis, CNRS
1361 route des Lucioles
06560 Valbonne, France}
\author{P. Vignolo}
\email{Patrizia.Vignolo@inln.cnrs.fr}
\affiliation{Institut Non Lin\'eaire de Nice UMR7335,
Universit\'e de Nice-Sophia Antipolis, CNRS
1361 route des Lucioles
06560 Valbonne, France}
\begin{abstract}
  We introduce a two-dimensional short-range correlated disorder that
  is the natural generalization of the well-known one-dimensional dual
  random dimer model [Phys. Rev. Lett \textbf{65}, 88 (1990)].  We
  demonstrate that, as in one dimension, this model induces a
  localization-delocalization transition in the single-particle
  spectrum. Moreover we show that the effect of such a disorder on a
  weakly-interacting boson gas is to enhance the condensate spatial
  homogeneity and delocalisation, and to increase the condensate
  fraction around an effective resonance of the two-dimensional dual
  dimers.  This study proves that short-range correlations of a
  disordered potential can enhance the quantum coherence of a
  weakly-interacting many-body system.
\end{abstract}

% 67.85.Hj 	Bose-Einstein condensates in optical potentials
%71.23.An=(Electronic structure of disordered solids.) 
%Theories and models; localized states 
\maketitle

\section{\label{sec:intro}Introduction}
The presence of impurities usually deeply modify the nature of the
spectrum of a quantum system, and thus its coherence and transport
properties.  In the absence of interactions, if the impurity
distribution is completely random, all states of the spectrum are
exponentially localized in dimensions one (1D) and two (2D), while a
mobility edge exists in dimensions three
(3D)\cite{Ande58,Ande79,Akkermans2007}.  If the impurity positions are
correlated, as for instance if it exists a minimum distance between
the impurities \cite{schaff10,Larcher2013}, some delocalized states
can appear in the spectrum. This was demonstrated in 1D in the context
of the Random Dimer Model (RDM) and of the Dual Random Dimer Model
(DRDM) \cite{Phil90}. In 1D, the effects of correlated disorder was
studied in different physical contexts (see for instance
\cite{deMour98,Izrailev1999,Tessieri2002,Kuhl08,Luga09}).  In 2D, the
effect of correlations is almost unexplored, except for the case of a
speckle potential \cite{Kuhn07}, and for the case of
pseudo-2D random dimer lattices with separable
dimensions \cite{Naether2015}. Correlations in speckle potentials
may mimic the presence of a mobility edge \cite{Luga09}, but
in the thermodynamic limit all states are localized \cite{Kuhn07}.
Random dimers introduce a set of delocalized states in 
pseudo-2D lattices \cite{Naether2015} as in 1D \cite{Phil90}. 
From a statistical point of view, the main difference between these two models
is the decay of the correlation function that is algebraic for the first
and exponential for the second. This ``short-range'' feature of the 
random dimer model is at the basis of the delocalization mechanism.

In interacting systems, the presence of disordered impurities gives
rise to a remarkable richness of phenomena. For instance, the
condensate and the superfluid fraction are modified by the presence of
the disorder \cite{Astra2002,Buonsante2009}, and this can shift the
onset of superfluidity \cite{Pilati2010,Plisson2011,Allard2012}, and,
on lattice systems, can induce exotic phases such as the Bose glass
\cite{Fisher1989}.

In this work we study the effect of a short-range correlated disorder on 
a Bose gas confined on a 2D square lattice. First we introduce a 2D
generalization of the DRDM (2D-DRDM).  In such a model, impurities
cannot be first neighbours and each impurity also modifies the hopping
with its nearest neighbor sites.  Using a decimation and
renormalization procedure \cite{Farchioni1992a}, we show that, in the
non-interacting regime, it exists a resonance energy at which the
structured impurity is transparent and the states around this energy
are delocalized.  It is remarkable that this resonance energy does not
depend on the system dimensionality and it is the same as the DRDM
in 1D~\cite{Phil90,schaff10}.  Then, we consider the case of a weakly
interacting Bose gas confined on such a potential.  Within a
Gutzwiller approach, we show that the effect of the 2D-DRDM is to
drive the homogeneity of the ground state. The disorder induces a
non-monotonic behavior of the condensate spatial delocalization and
of the condensate fraction as a function of the disorder strength, and
enhances both in correspondence of the resonance energy of 2D-DRDM
single-particle Hamiltonian. We show that the dependence of such
quantities on the interaction strength can be explained by including
the effect of the healing length in the resonance condition
discussion.

The manuscript is organized as follows. In Sec. \ref{sec:model}, we
introduce the 2D-DRDM potential and we demonstrate its single-particle
delocalization properties in the region of the spectrum around the
resonance energy.  The effect of such a potential on a
weakly-interacting Bose gas is studied in Sec. \ref{sec:results},
where we also introduce a suitable inverse participation ratio for our
many-body system and study it for the case of the 2D-DRDM potential
and for an uncorrelated random disorder. Moreover, we compute the
density distribution and the condensate fraction as functions of the
disorder strength.  Our concluding remarks in Sec. \ref{sec:concl}
complete this work.

\section{\label{sec:model} The DRDM in two dimensions}
We consider the tight-binding single-particle Hamiltonian
\begin{equation}
  H=-\sum_{\langle ij \rangle}t_{ij}\left(|i\rangle\langle j|+ |j\rangle\langle i|\right)+
  \sum_{i=1}^N \varepsilon_i |i\rangle\langle i|
\end{equation}
where $\varepsilon_i$ are the on-site energies, $t_{ij}$ the 
first neighbor hopping terms, $N$ the number of sites and
$\langle ij \rangle$ denotes the sum over first neighbor sites.

\begin{figure}
\begin{center}
\includegraphics[width=0.9\linewidth]{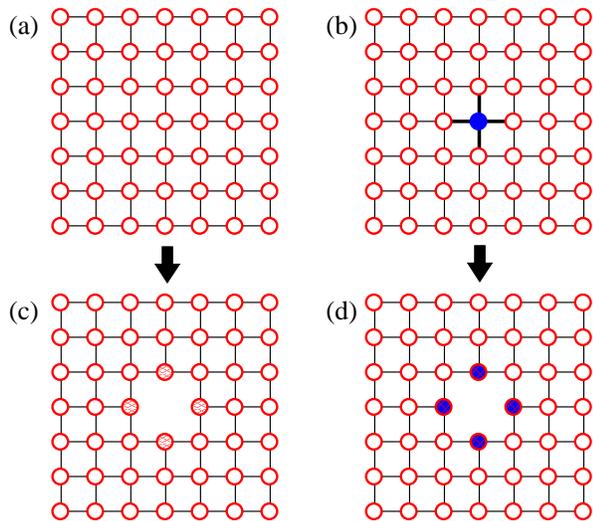}
\caption{\label{fig1} Schematic representation of (a) the unperturbed
  Hamiltonian; (b) the Hamiltonian in the presence of a single
  impurity; (c) the effective Hamiltonian after decimation of the site
  0 in the Hamiltonian (a); (d) the effective Hamiltonian after
  decimation of the site 0 in the Hamiltonian (b).}
\end{center}
\end{figure}

We focus on a 2D square lattice of linear dimension $L$ ($N=L^2$
lattice sites), and compare the ordered lattice with $\varepsilon_i=0$
and $t_{ij}=t$ $\forall$ $\langle i j\rangle$, as schematized in
Fig.~\ref{fig1}(a) with a lattice where we introduce an impurity at
the site 0, $\varepsilon_0=\Delta$ that modifies the hopping terms
involving this site, $t_{0,j}=t'$ [Fig.~\ref{fig1}(b)].
\subsection{Effect of correlations in the single-particle spectrum}
With the aim of understanding the effect of the impurity, we consider
the Green's function $G_{AA}(E)=\langle A|(E-H)^{-1}|A\rangle$ projected on 
the subspace $A$, including all sites except the site $0$ 
with coordinates $(0,0)$.
Using a decimation and renormalization technique \cite{Farchioni1992a}, 
it can be shown that
\begin{equation}
G_{AA}(E)=(E-H_{eff})^{-1}
\end{equation}
with
\begin{equation}
H_{eff}=\left\{\begin{array}{ll}
H_{AA}+\dfrac{t_{0,j}^2}{E-\varepsilon_0}&\phantom{bla}{\rm if}\;j\;{\rm is\;a\;first\!-\!neighbour}\\[-2mm]
&\phantom{bla}{\rm site\;of\;the\;site\;}0\\[1mm]
H_{AA}&\phantom{bla}{\rm elsewhere}
\end{array}\right.
\end{equation}
where $H_{AA}=\langle A|H|A\rangle$. The effective Hamiltonian for the
unperturbed case in Fig. \ref{fig1} (a), is schematically illustrated
in Fig. \ref{fig1} (c); whereas the effective Hamiltonian for the case
with a single impurity in Fig. \ref{fig1} (b), is illustrated in Fig.\
\ref{fig1} (d).  The subspace $A$ does not ``feel'' the presence of
the impurity if $G_{AA}$ ($H_{eff}$) remains the same in the absence
or in the presence of the impurity, namely if
\begin{equation}
\dfrac{t^2}{E}=\dfrac{(t')^2}{E-\Delta}.
\label{condition}
\end{equation} 
The condition (\ref{condition}) is satisfied if
$E=E_{res}=-\dfrac{\Delta}{(t'/t)^2-1}$. If $E_{res}$ is an allowed energy 
of the system, namely if $-4t<E_{res}<4t$, at $E=E_{res}$ the impurity
will not affect the eigenstate at this energy (in the subspace $A$).

\begin{figure}
\begin{center}
\includegraphics[width=0.8\linewidth]{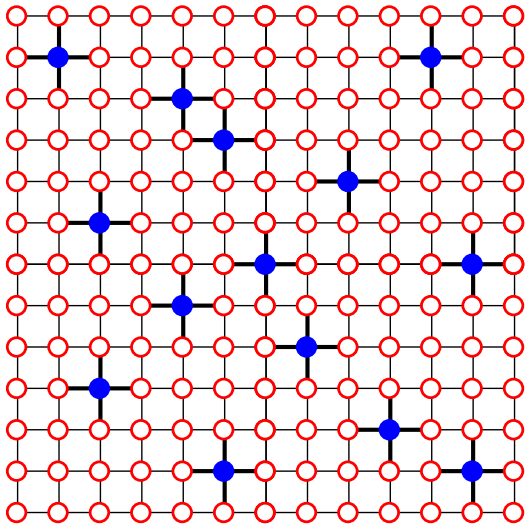}
\caption{\label{fig2} Schematic representation of the 2D DRDM.}
\end{center}
\end{figure}

If we add other impurities in the system, as the one in
Fig. \ref{fig1} (b), with the supplementary condition that on-site
impurities cannot occupy first neighbor sites (Fig. \ref{fig2}), one
can repeat the same argument as above, properly redefining the
subspace $A$, and one obtains exactly the same condition
(\ref{condition}) imposing that {\it all} the $N_{imp}$ impurities do
not perturb the system (the subspace $A$).  Thus at $E=E_{res}$, the
impurities are transparent as in the 1D DRDM~\cite{Phil90}.  Indeed,
with this procedure, we are defining a 2D-DRDM, where at each
``isolated'' impurity correspond a structure of 4 hopping terms
forming a cross, as shown in Fig.~\ref{fig2}. Let us remark that this
definition of the model provides the same condition (\ref{condition})
independently from the dimensionality of the system
\cite{Phil90,schaff10}. However our model is fully 2D and the Hamiltonian
cannot be mapped onto two 1D DRDM as opposed to Ref. \cite{Naether2015}. 

\begin{figure*}
\begin{center}
\begin{tabular}{cc}
\includegraphics[width=0.45\linewidth]{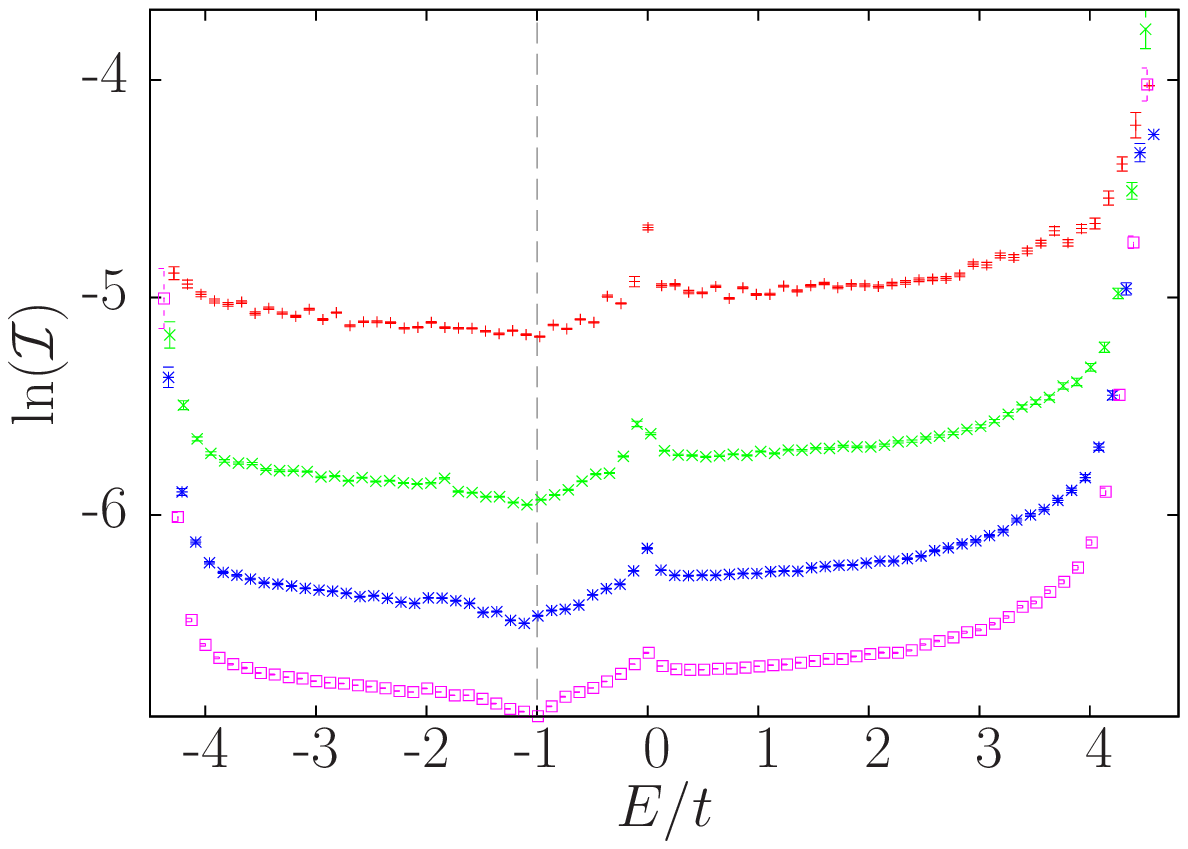} &
\includegraphics[width=0.45\linewidth]{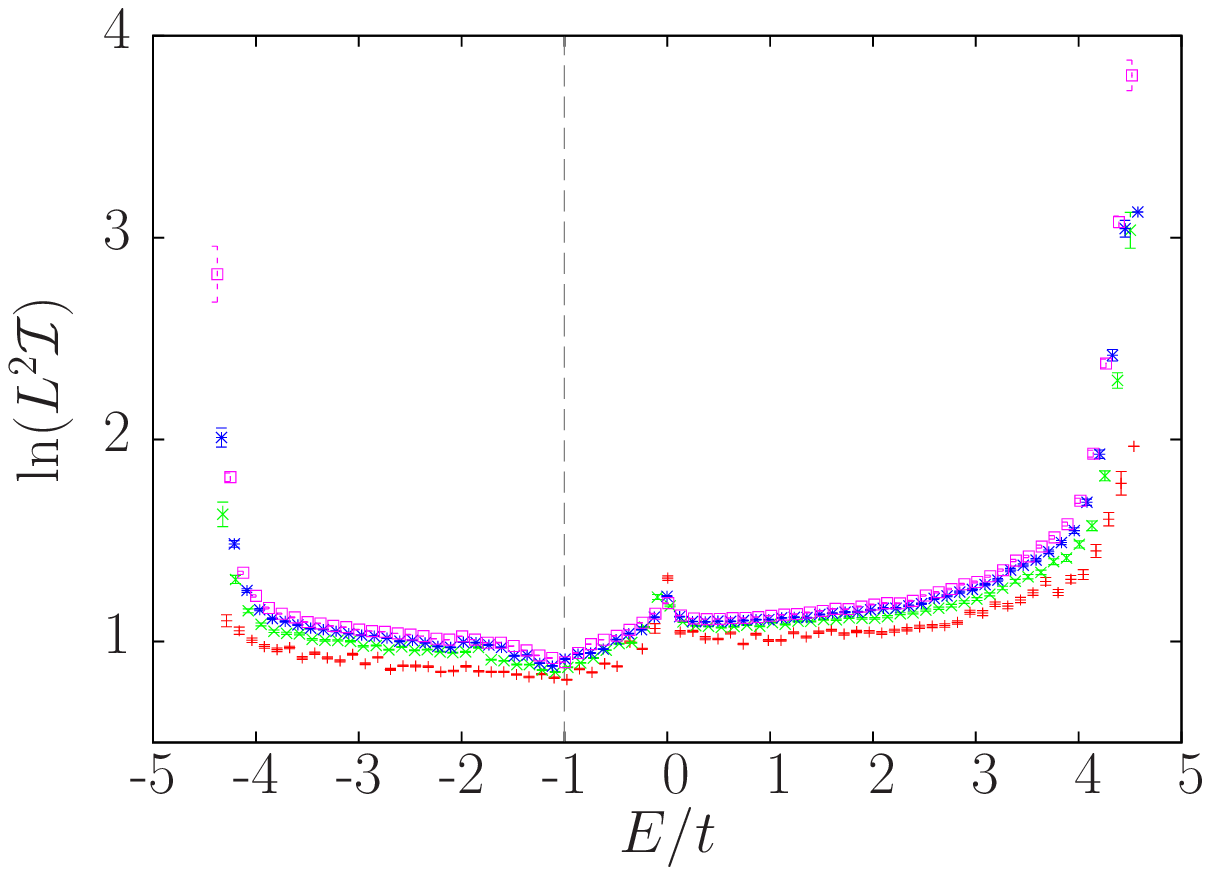} \\
\includegraphics[width=0.45\linewidth]{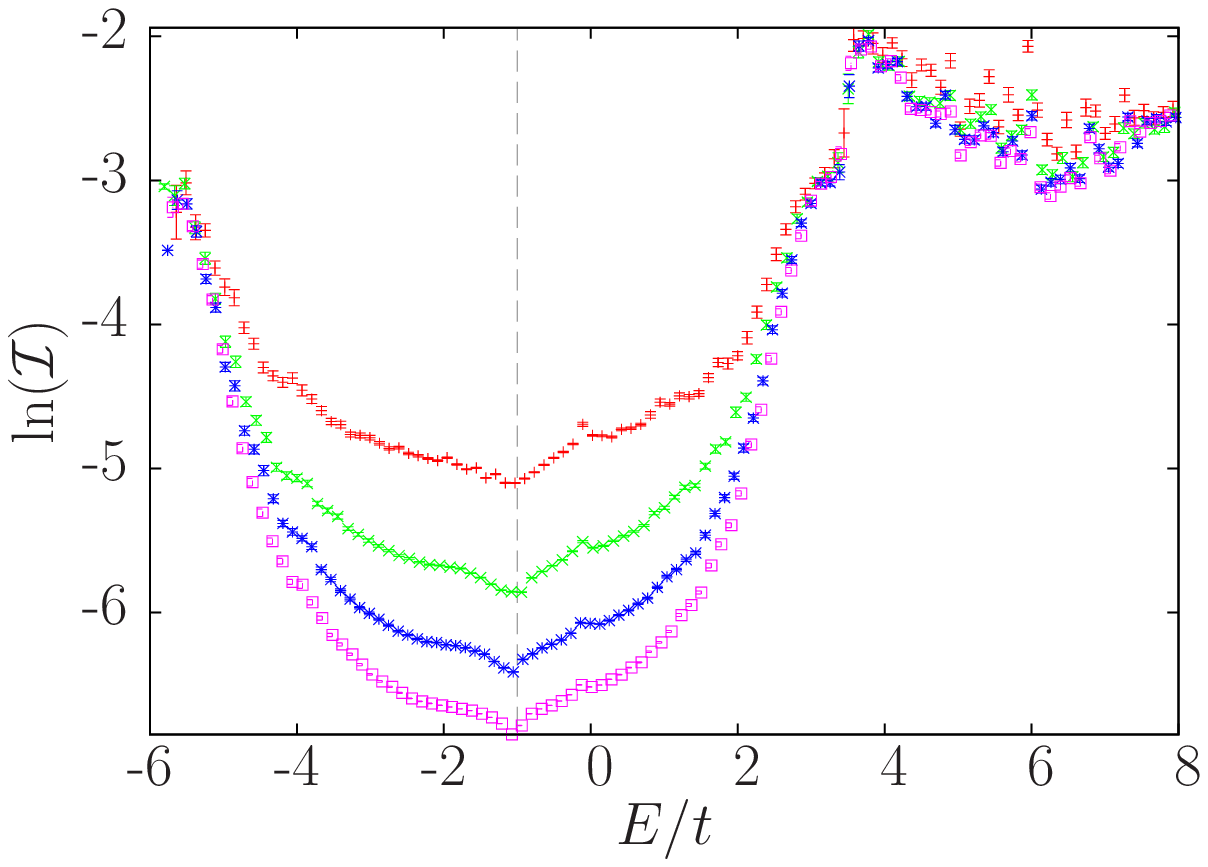}&
\includegraphics[width=0.45\linewidth]{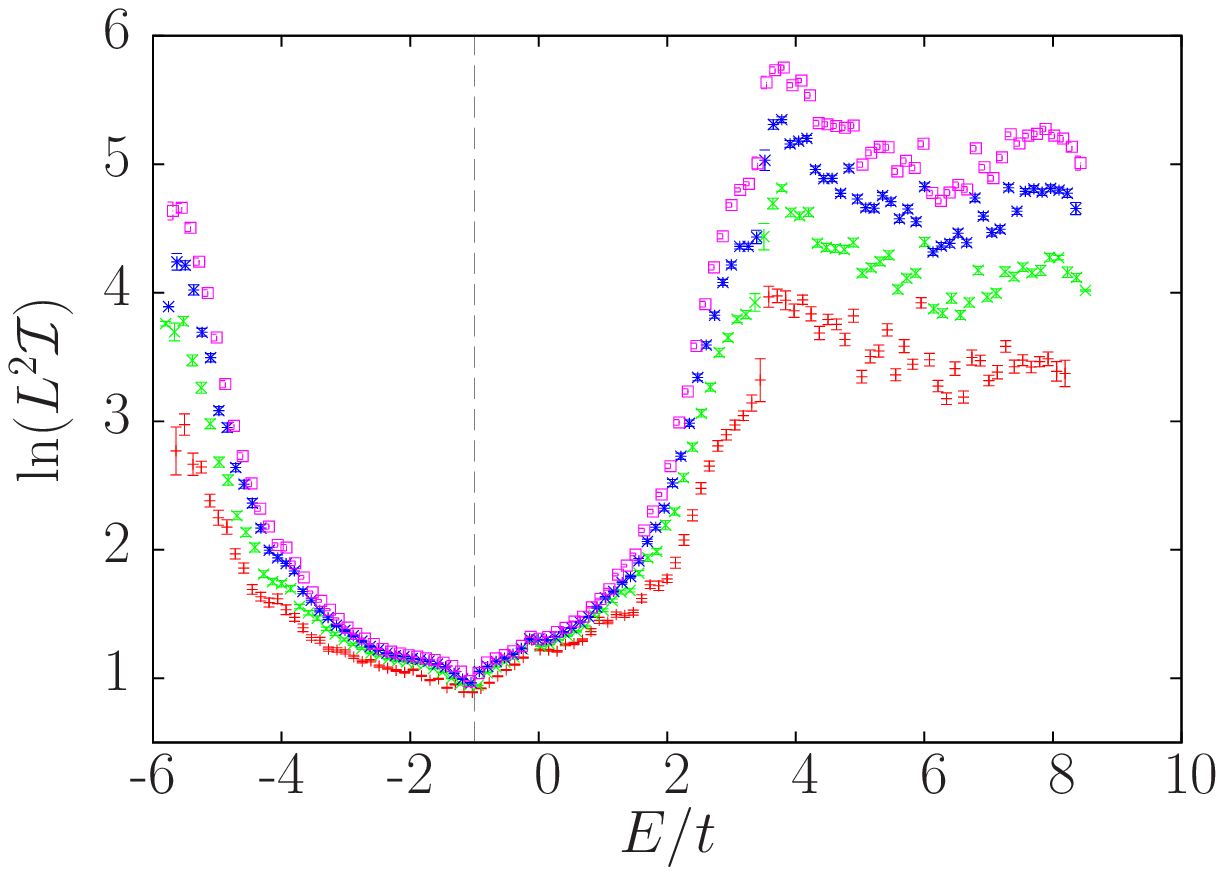} \\
\includegraphics[width=0.45\linewidth]{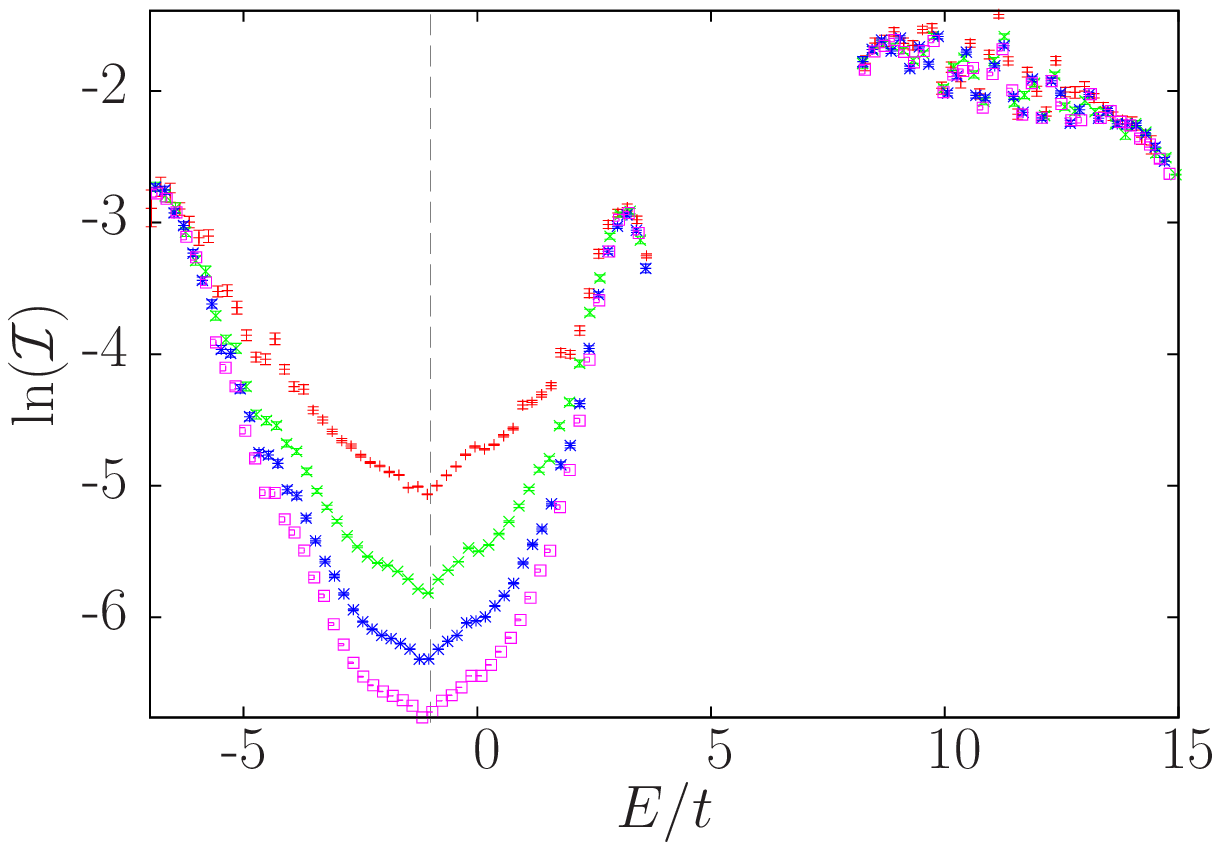}&
\includegraphics[width=0.45\linewidth]{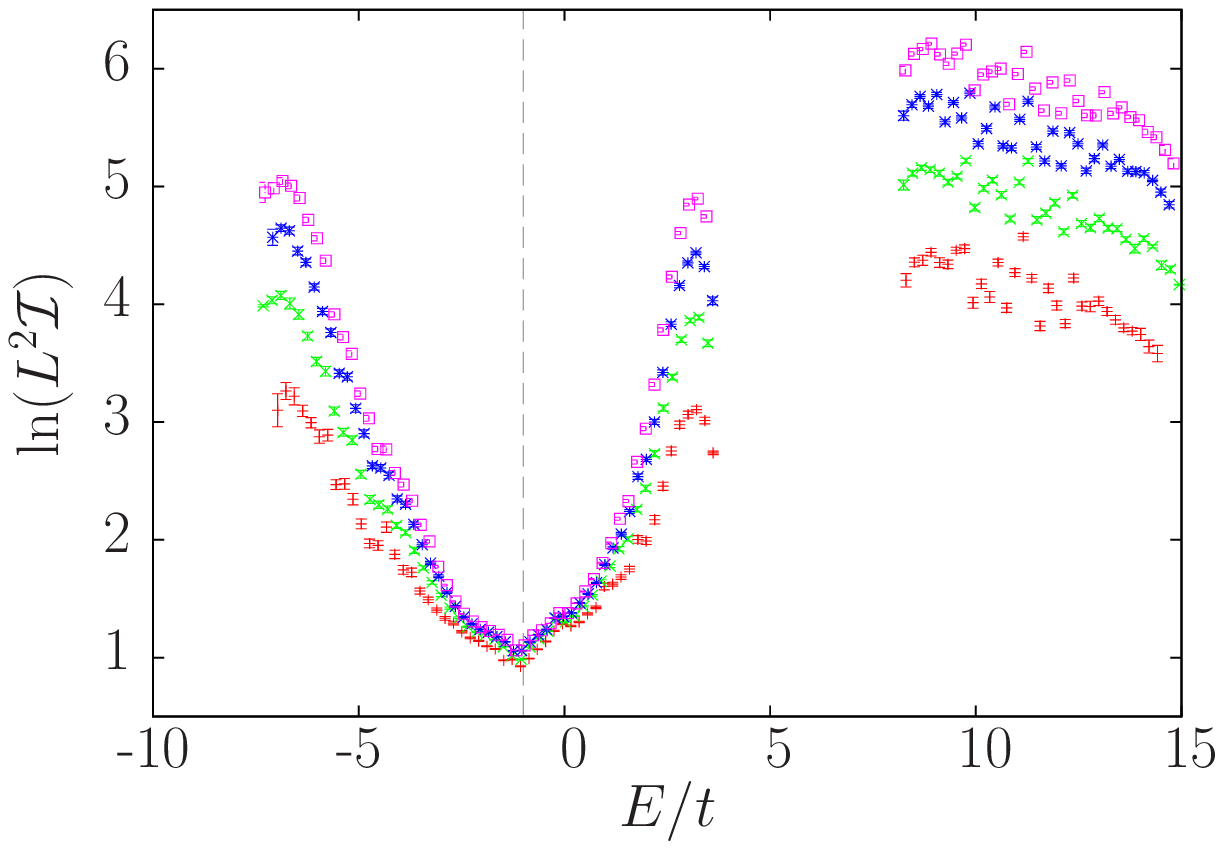}
\end{tabular}
\end{center}
\caption{\label{fig3} (Color online) Inverse participation ratio
  ($\ln(\mathcal{I})$ left column and $\ln(\mathcal{I}\,L^2)$ right
  column) as a function of the energy $E$ in units of $t$. The plots
  in the first row correspond to $\Delta/t=0.44$ and $t'/t=1.2$; in
  the second row correspond to $\Delta/t=3$ and $t'/t=2$; and in the
  the third row to $\Delta/t=8$ and $t'/t=3$. The different curves in
  each plot correspond to different system sizes: $L=20$ (red), 30
  (green), 40 (blue) and 50 (magenta).  Each curve correspond to
  $N_{imp}/N\simeq 0.15$ and to an average over 50 configurations.
  The data are binned in 80 (first row) and 110 bins (second and third
  rows). The vertical dashed lines indicate $E_{res}$.}
\end{figure*}

With the aim of analyzing the localization properties of this model,
we consider the Inverse Participation Ratio (IPR)
\begin{equation}
  \mathcal{I}(E)=\left\langle\dfrac{\sum_i |\psi_i(E)|^4}{(\sum_i |\psi_i(E)|^2)^2}\right\rangle.
\label{eq:ipr}
\end{equation}
The symbol $\langle\dots\rangle$ denotes the average over different
disorder configurations, and $\psi_i(E)$ the wavefunction on site $i$ and 
at energy $E$.  If $E_\alpha$ is an eigenvalue of the system
and $\psi_i(E=E_\alpha)$ is an extended state, then $\mathcal{I}$
decreases as a function of $L$.  On the other side, if
$\psi_i(E_\alpha)$ is a localized state, then $\mathcal{I}$ does not
depend on $L$ (if $L$ is larger than the localization length).  In
Fig. \ref{fig3} we show the behavior of $\ln(\mathcal{I})$ and
$\ln(\mathcal{I}\,L^2)$ (column left and right respectively), for the
Hamiltonian illustrated in Fig. \ref{fig2}.

We consider three set of parameters, (i) $\Delta/t=0.44$ and
$t'/t=1.2$, (ii) $\Delta/t=3$ and $t'/t=2$, and (iii) $\Delta/t=8$ and
$t'/t=3$ that give the same resonance energy, $E_{res}/t=-1$.  In all
the three cases, the curves $\ln(\mathcal{I}(E)\,L^2)$ collapse around
$E=E_{res}$ meaning that the states are delocalized in this energy
region. Moreover, due to the large strength of the disorder,
the spectrum varies considerably for the cases (ii) and (iii),
and an energy gap appears in (iii).

\begin{figure}
  \begin{center}
    \includegraphics[width=0.9\linewidth]{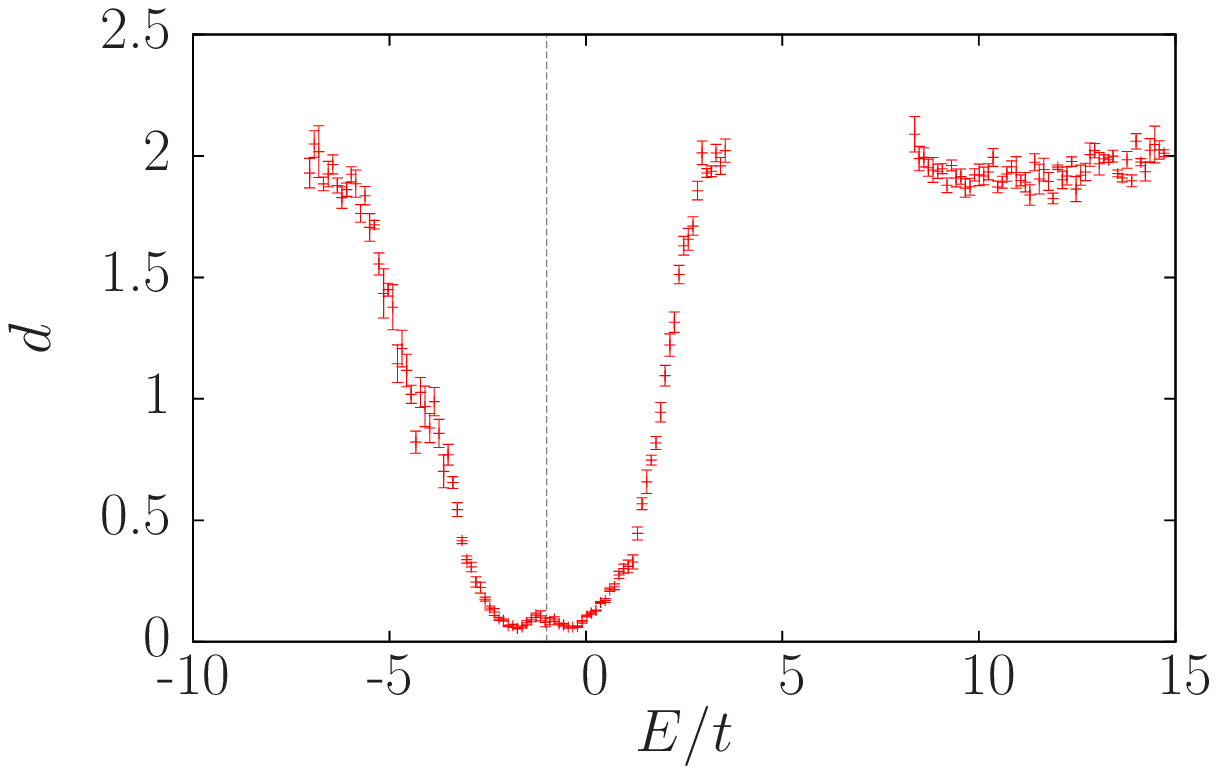}    
    \caption{The exponent of Eq.~(\ref{eq:asymPatty}) as a function of
      the energy for $\Delta/t=8$ and $t'/t=3$. The exponent has been
      obtained using calculations for lattice-linear dimensions
      $L=40,50,60,70,80,90,100$ averaged over 20 realizations,
      the error bars correspond to the standard deviation of the
      fit of the data to Eq.\ (\ref{eq:asymPatty}).  The vertical
      dashed line indicates $E_{res}$.}
\label{fig:esponenti}
  \end{center}
\end{figure}

The inverse participation ratio, Eq.~(\ref{eq:ipr}), in two dimensions has the
following asymptotic behavior~\cite{murphy2011}
\begin{equation}
\lim_{L\rightarrow\infty}\mathcal{I}(E)=\left\{%
  \begin{matrix}
    1/L^2 & \text{(extended states)} \\
    \text{const.} & \text{(localized states)}\,. \\
  \end{matrix}\right.
\end{equation}
Thus, the asymptotic behavior of the function $\mathcal{I}(E)\,L^2$ is 
\begin{equation}
  \lim_{L\rightarrow\infty}\mathcal{I}(E)\,L^2= L^d\,,
  \label{eq:asymPatty}
\end{equation}
with $d=2$ for localized states, and $d=0$ for extended states. In
Fig.~\ref{fig:esponenti} we have analyzed the exponent $d$ as a
function of the energy for the set of parameters (iii). We observe a
high-energy band of localized states that has been created by the
disorder; the original (without noise) band has been distorted, and
the states at its boundaries are localized. The center of the band,
around $E_{res}$, is mainly constituted of extended states.
The width of the feature around $E_{res}$
corresponds to the width of the resonance dip of the inverse
participation ratio at this energy value (Fig. \ref{fig3}).

These results confirm that our 2D extension of the DRDM introduced by
Dunlap and collaborators in Ref.~\cite{Phil90} for 1D systems
introduces a set of delocalized states even at higher dimensions.

\section{\label{sec:results} Effects of the interactions}
We now consider the case of weakly interacting bosons in the presence
of the potential defined in Sec. \ref{sec:model}.  This system is
described by the Bose-Hubbard Hamiltonian in the grand canonical
ensemble
\begin{equation}
  H_{BH}=-\sum_{\langle ij \rangle}t_{ij}(\hat a_i^\dagger\hat a_j+\hat a_j^\dagger \hat a_i)-
  \sum_i(\mu-\varepsilon_i)\hat{n}_i+\dfrac{U}{2}\sum_i \hat{n}_i(\hat{n}_i-1)
\label{eq:BHGW}
\end{equation} 
where $\hat a_i^{\dagger}$ is the creation operator defined at the
lattice site $i$, $\hat{n}_i=\hat a_i^\dagger\hat a_i$, $U$ the
interparticle on-site interaction strength, and $\mu$ denotes the
chemical potential fixing the average number of bosons.

We use a Gutzwiller approach to find the ground state wavefunction for
a given set of parameters and average number of particles. The
Gutzwiller ansatz is given by the site product wavefunction in the
occupation number representation
\begin{equation}
|\Phi_{\mathrm{GS}}\rangle = \prod_i^{L\times L} \sum_{n_i} f_i(n_i)|n_i\rangle, 
\end{equation}
where $f_i(n_i)$ are the probability amplitudes of finding $n_i$
particles on site $i$. The ansatz provides an interpolating
approximation correctly describing both the Bose-condensed and
Mott-insulating phases for low and high $U$, respectively, in
dimensions larger than one.  In addition, the approximation becomes
exact for all $U$ in the limit of infinite dimensions
\cite{Rokhsar1991,Metzner1990}.

We minimize the average energy given by Hamiltonian (\ref{eq:BHGW}) as
a function of the set of amplitudes $f_i(n_i)$ with the normalization
and average number of particle constraint for at least 30 disorder
realizations for each set of parameters. The minimization is done
using standard conjugate-gradient and/or Broyden-Fisher techniques
\cite{Press2007} which provides reasonable performance for moderate
lattice sizes.

\subsection{Characterization of the condensate delocalization}   
To quantify the extent of delocalization of the ground state
$|\Phi_{\mathrm{GS}}\rangle$ in the interacting regime, we decompose
it onto the localized basis $|\psi_i\rangle$,
$|\Phi_{\mathrm{GS}}\rangle=\sum_i c_i |\psi_i\rangle$, representing
the distribution of a homogeneous condensate with average density $n$
on the lattice \cite{Dukesz2009}. We define the many-body ground-state
IPR $\mathcal{I}_{\mathrm{GS}}$ with respect to this basis as
\begin{equation}
  \mathcal{I}_{\mathrm{GS}}=\left\langle\dfrac{\sum\limits_{i=1}^{N}c_i^4}{(\sum\limits_{i=1}^{N}c_i^2)^2}\right\rangle.
\label{eq:IPRGS}
\end{equation}
$\mathcal{I}_{\mathrm{GS}}$ measures the homogeneity of the ground
state in the condensation regime: the smaller
$\mathcal{I}_{\mathrm{GS}}$ the more spatially delocalized is the
condensate.

\begin{figure}
\includegraphics[width=0.9\linewidth]{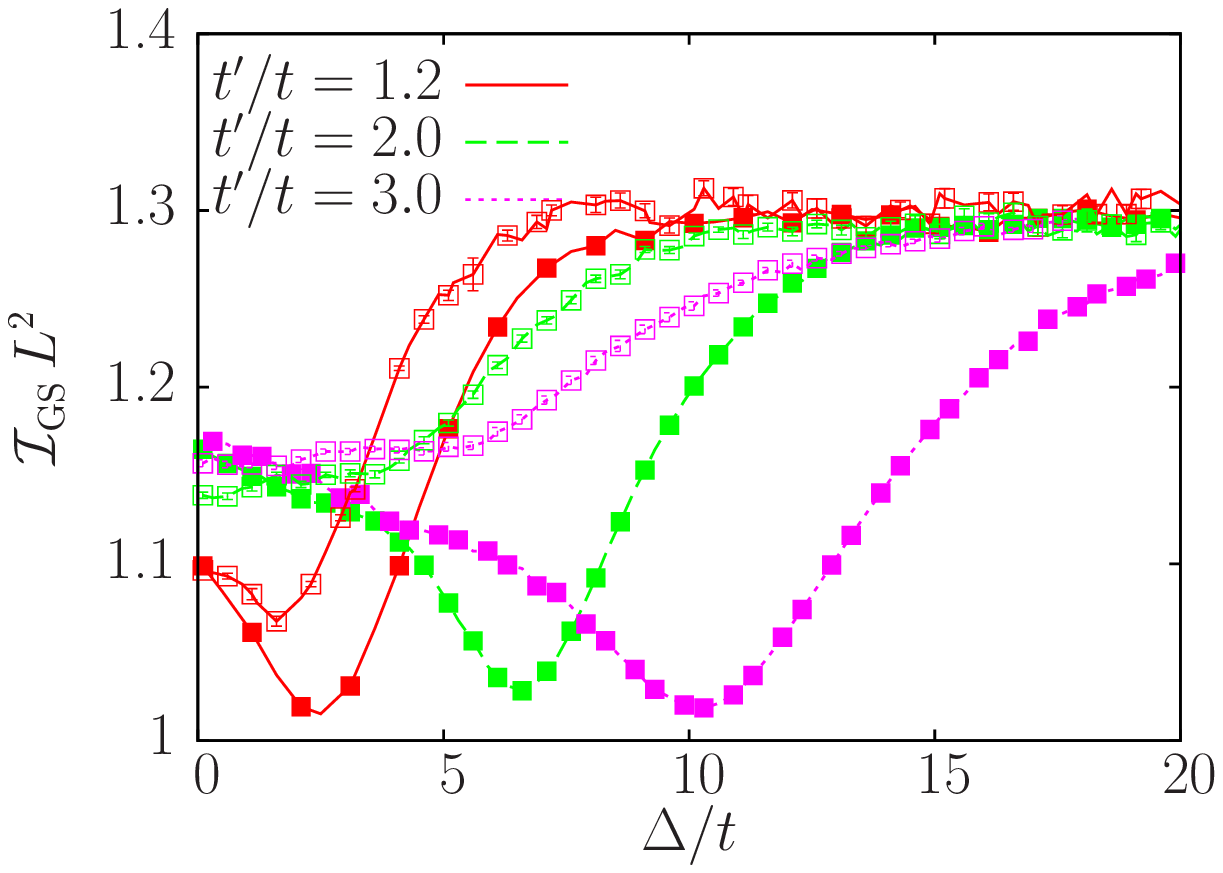}
\caption{\label{fig:IGS_L20}(Color online)
$\mathcal{I}_{\mathrm{GS}}L^2$ as a function of $\Delta/t$
for $L=20$, $U/t=10^{-2}$ and $n=20$ particles per site. 
The different curves correspond to different values of $t'$ as indicated 
in the figure.
The filled symbols correspond to the 2D-DRDM potential and the empty 
symbols correspond to the UN-RAND potential.}
\end{figure}

In Fig.\ \ref{fig:IGS_L20} we show the behavior of
$\mathcal{I}_{\mathrm{GS}}$ as a function of $\Delta$, by fixing
$L=20$, $U/t=10^{-2}$ and $n=20$, for several values of $t'$.
We compare the case of $22\%$ of correlated impurities $N_{imp}$ with
the one of the same percentage of uncorrelated impurities, where there
is no restriction for the position distribution of the on-site
impurities $\Delta$ and no correlations between them and the
additional hopping $t'$ (UN-RAND). We note that due to the
correlations present in the 2D-DRDM the maximum percentage of allowed
impurities is $50\%$ (in this limit the system would be an ordered
checkerboard). We can observe that, in the case of the 2D-DRDM
potential, $\mathcal{I}_{\mathrm{GS}}$ has a minimum as a function of
$\Delta$, whose position depends on the value of $t'$.  This
non-monotonic behavior is a signature of the resonance induced by the
correlations of the disordered potential.  Indeed, it disappears for
the case of the UN-RAND potential and for large values of $t'$ (strong
disorder). The dip in the $\mathcal{I}_{\mathrm{GS}}$ for the UN-RAND
potential and weak disorder ($t'/t=1.2$) indicates that some DRDM
impurities may still statistically appear, in the absence of
correlations. The effect of such impurities is not fully destroyed by
the other defects only if the strength of the disorder is weak.

\subsubsection{The resonance effect as a function of the interactions}
\begin{figure}
\includegraphics[width=0.90\linewidth]{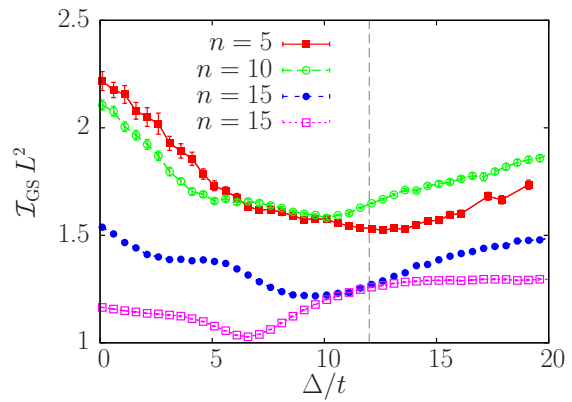}
\caption{\label{fig:IGS_L20_N}(Color online)
  $\mathcal{I}_{\mathrm{GS}}L^2$ as a function of $\Delta/t$ for
  $L=20$, $U=10^{-2}t$ and $t'/t=2$.  The different curves correspond
  to different values of the average density $n$ as indicated in the
  figure.  All the curves correspond to the 2D-DRDM potential.
  The vertical dashed line indicates the non-interacting resonance
condition given in Eq. (\ref{cond-pat}).}
\end{figure}

In the perturbative regime for negligible interactions, one would
expect that correlations modify the ground state if
$E_{res}=E_{\mathrm{GS}}$, $E_{\mathrm{GS}}$ being the ground-state
energy per particle, which corresponds to $\simeq -4t$ in the weak
disorder regime.  This condition, that can be written
\begin{equation}
\Delta=4t[(t'/t)^2-1],
\label{cond-pat}
\end{equation}
determines the location of the minimum of $\mathcal{I}_{\mathrm{GS}}$
at $\Delta/t=1.67$ for $t'/t=1.2$, $\Delta/t=12$ for $t'/t=2$, and
$\Delta/t=32$ for $t'/t=3$. However, in the limit of strong disorder, 
due to the interactions these
values strongly differ from those shown in
Fig.~\ref{fig:IGS_L20}. In fact, we calculate $\mathcal{I}_{\mathrm{GS}}$ for
smaller values of $n$ and verify that the minimum location of
$\mathcal{I}_{\mathrm{GS}}$ depends on $E_{res}$ and that the shift observed is
indeed an effect of the interactions. The results are illustrated in
Fig.~\ref{fig:IGS_L20_N}, where we focus on the case $t'/t=2$.
By decreasing the value of $n$, the minimum position $\Delta_{min}/t$
of $\mathcal{I}_{\mathrm{GS}}$ shifts from $6.5$ to about $12$ as
expected by the perturbative argument.  This shift can be understood
as follows. The interactions introduce the so-called healing length
$\xi=\sqrt{t/(2nU)}$ \cite{Bukov2014} that represents a coherence
length over which the system feels the effect of an impurity, or in
other words, the distance a site affects its neighborhood. For
$U/t=10^{-2}$ and $n$ from 20 to 5, the value of $\xi$ ranges
approximately from $1.5$ to 3 lattice spacing $\ell$, which shows that,
already for this $U$ value, the role of the interactions is important,
effectively reducing the coherence length.  To quantify this effect,
we can partition the system into independent boxes of dimension
$\xi\times\xi$, and use a mode-matching argument to determine their
ground states: the condensate is more homogeneous if the lowest
eigenvalue of each box is the same despite the presence of an
impurity.

\begin{figure}
\includegraphics[width=0.6\linewidth]{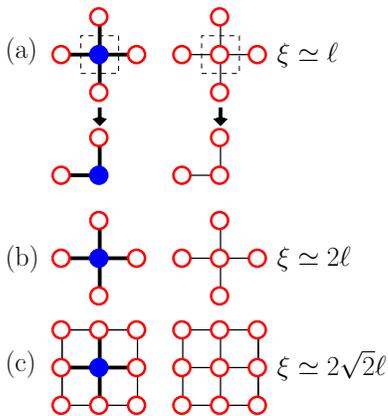}
\caption{\label{figboxes}(Color online)''Boxes'' of different sizes, 
in the presence and in the absence of an impurity.}
\end{figure}
Therefore, this mode-matching argument fixes the value of $\Delta$. 
  For the case $U/t=10^{-2}$ and $n=20$,
$\xi\simeq 1.6\ell$, and this gives $4.24<\Delta/t<6$ while for $n=5$,
$\xi\simeq 3.2\ell$, and we expect to find $8.4<\Delta/t<12$, in good
agreement with the results showed in Fig.~\ref{fig:IGS_L20_N}.  
Namely, the larger is $\xi$, the better we recover the non-interacting
condition Eq. (\ref{cond-pat}). This effect is summarized in Table
\ref{tab:Uloco}.
\begin{table}
  \caption{\label{tab:Uloco}Effective linear dimensions $\xi$ and positions of
the expected resonance $\Delta$ for the weakly interacting bosons in the
2D-DRDM. }
\begin{center}
\tabcolsep=6pt
\begin{tabular}{ccc}
\hline
\hline
$ \xi$ &Fig. & $\Delta$  
  \\ \hline  \\
$\ell$ &\ref{figboxes}(a)&$\sqrt{2}t\,[(t'/t)^2-1]$\\[5pt]
$2\ell$& 
\ref{figboxes}(b)&\; $2t\,\,[(t'/t)^2-1]$\\[5pt]
$2\sqrt{2}\ell$& 
 \ref{figboxes}(c)& $2\sqrt{2}t[(t'/t)^2-1]$ \\[5pt]
\hline \hline
\end{tabular}
\end{center}
\end{table}

We remark that this mode-matching condition is equivalent to match the
resonance energy $E_{res}$ with the lowest eigenvalue of the
unperturbed system of size $\xi\times\xi$.  These simple arguments,
allow us to understand the shift of $\Delta$ as a function of the
interaction energy $Un$ and the role of the structured impurities 
in the presence of the interactions.

\subsubsection{The resonance effect as a function of the system size}

\begin{figure}
\includegraphics[width=0.93\linewidth]{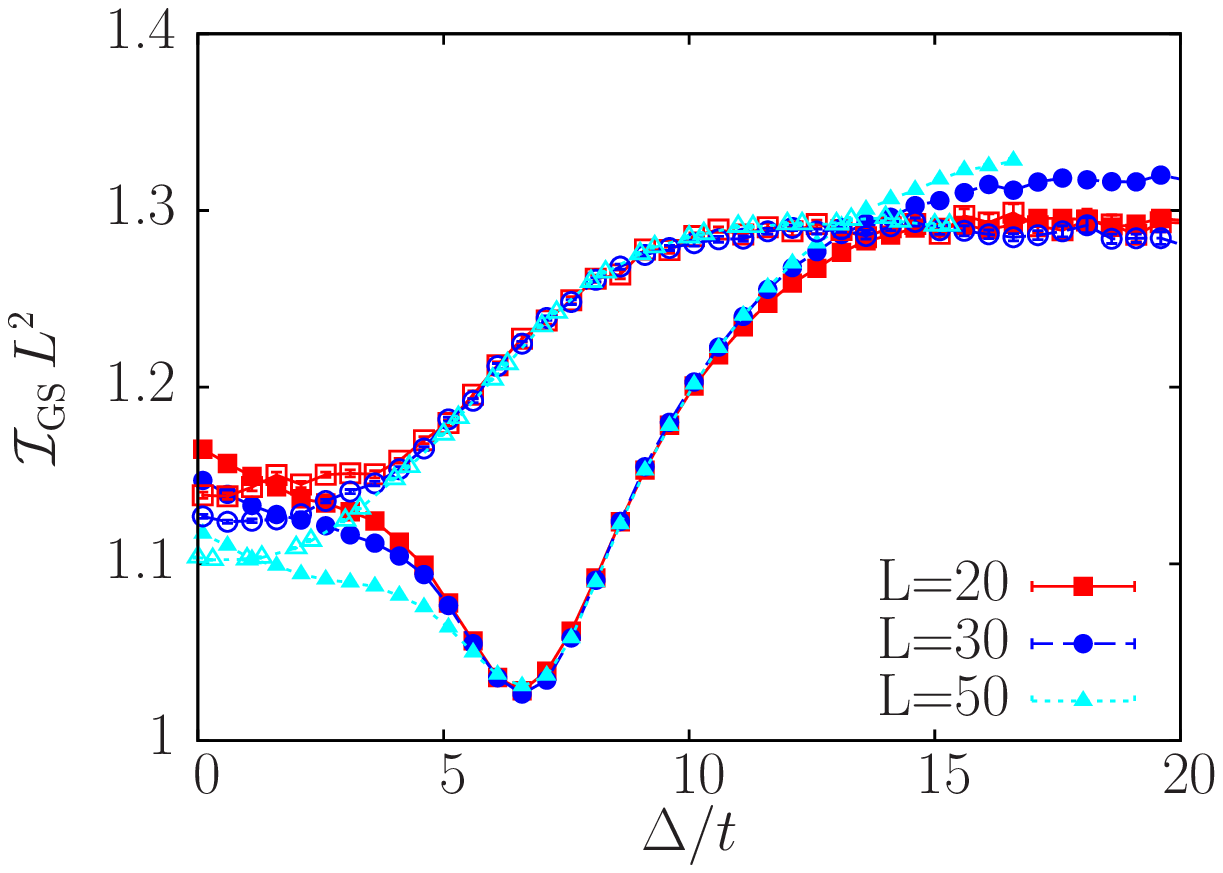}
\caption{\label{fig:IGS_L} (Color online)
  $\mathcal{I}_{\mathrm{GS}}L^2$ as a function of $\Delta/t$ for
  $t'/t=2$, $U/t=10^{-2}$ and $n=20$ particles per site. The different
  curves correspond to different values of $L$ as indicated in the
  figure.  The filled symbols correspond to the 2D-DRDM potential and
  the empty symbols correspond to the UN-RAND potential.}
\end{figure}

We study the scaling behavior of $\mathcal{I}_{\mathrm{GS}}L^2$ with respect
to $L$.  Analogously to the case of the single-particle IPR
$\mathcal{I}(E)$ [see Eq. (\ref{eq:asymPatty})], we expect that
\begin{equation}
  \lim_{L\rightarrow\infty}\mathcal{I}_{\mathrm{GS}}\,L^2= L^d\,,
  \label{eq:asymPatty2}
\end{equation}
with $d=2$ for a condensate localized on few sites, and $d=0$ 
for a homogeneous extended condensate.
The behavior of $\mathcal{I}_{\mathrm{GS}}L^2$ for different values of $L$
is shown in Fig. \ref{fig:IGS_L}.
We observe that the minima, corresponding to different system sizes,
collapse all together, meaning that the ground state corresponds to a
spatial homogeneous condensate in the parameter regime where the
correlations are dominant.  At lower values of $\Delta$,
$\mathcal{I}_{\mathrm{GS}}L^2$ scales as $L^{-\epsilon}$, and larger
values of $\Delta$, $\mathcal{I}_{\mathrm{GS}}L^2$ scales as
$L^{\epsilon'}$, with $\epsilon$ and $\epsilon'$ $>0$.  This sort of
``super-delocalization'', in the low $\Delta$ region, is determined by
the large value of $t'$ that compensates, in the structured impurities,
the effect of the site defect.  Indeed, we observe
an analogous behavior for the UN-RAND potential. For such a
potential, where the effect of $t'$ is no more dominant, all the
curves collapse together. Thus we expect that in this region the
effect of the uncorrelated impurities on the ground state
density distribution does not depend on the system size.

\subsection{Condensate delocalization and condensate fraction}
\begin{figure}
\centering
\includegraphics[width=0.9\linewidth]{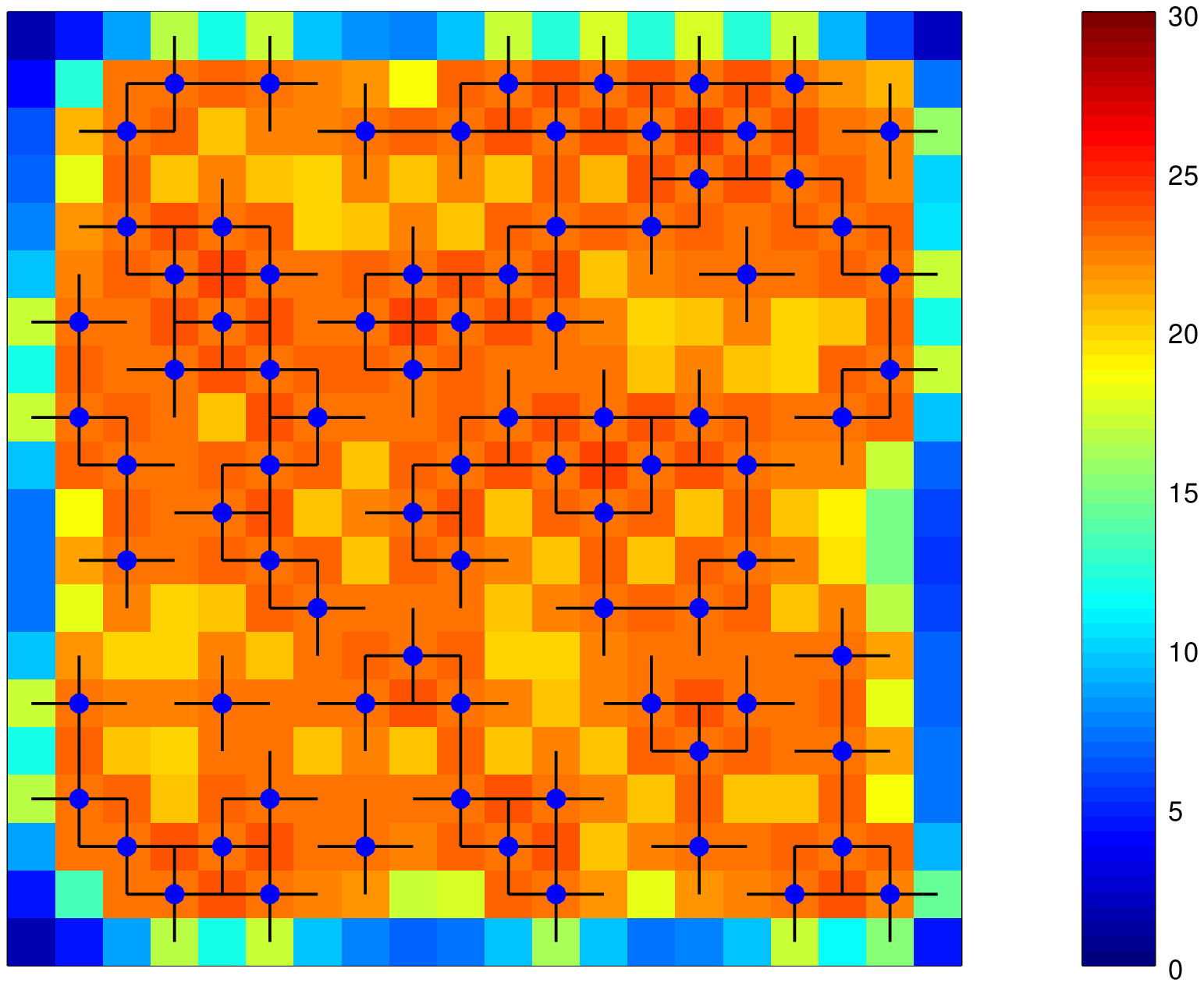}\\
\includegraphics[width=0.9\linewidth]{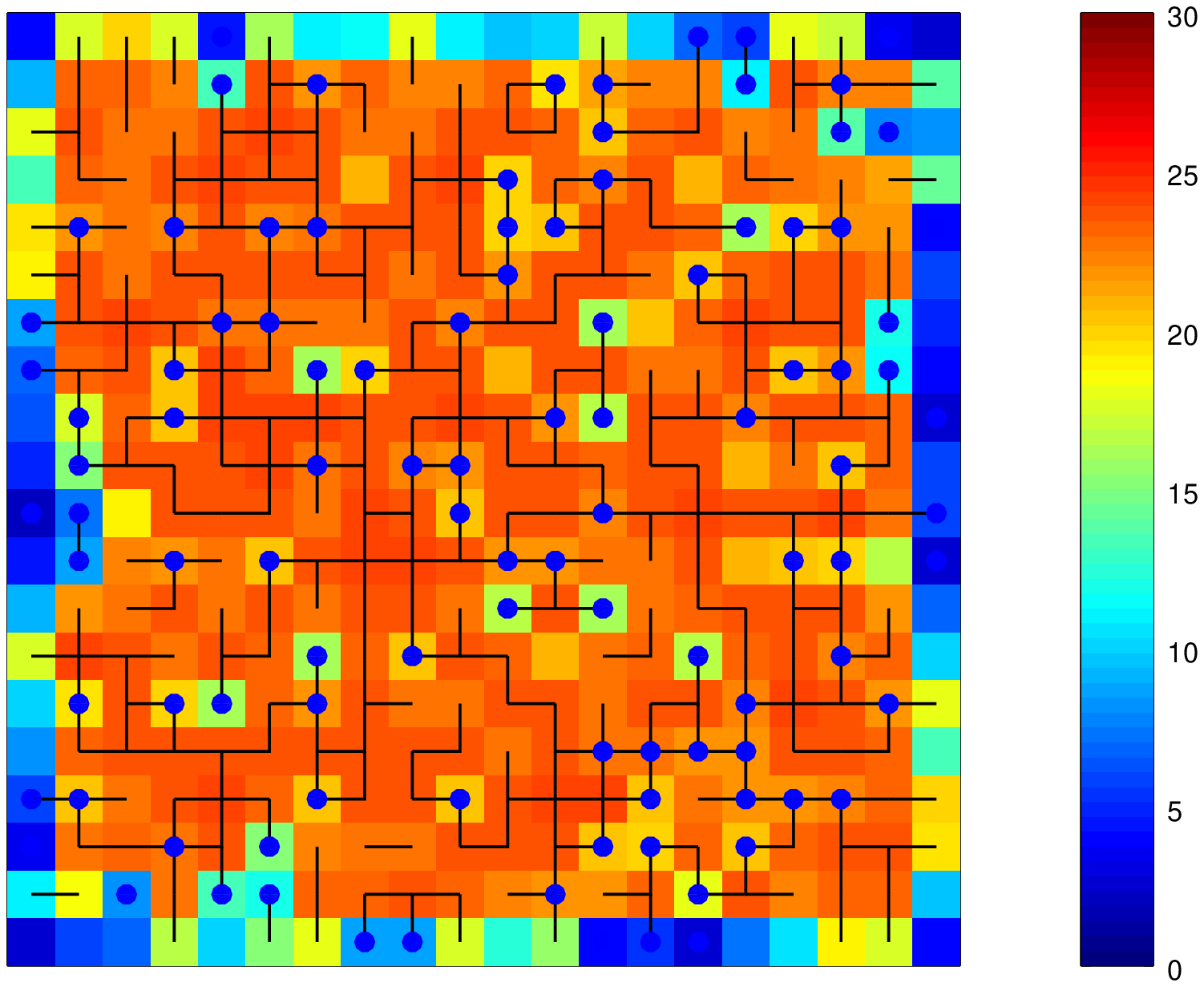}
\caption{\label{fig:denlatt2.1}(Color online) Lattice density plots
  together with site and bond impurities locations for $t'/t=2$,
  $\Delta/t\simeq2$ and DRDM disorder (top) and UN-RAND (bottom).}
\end{figure}

\begin{figure}
\centering
\includegraphics[width=0.9\linewidth]{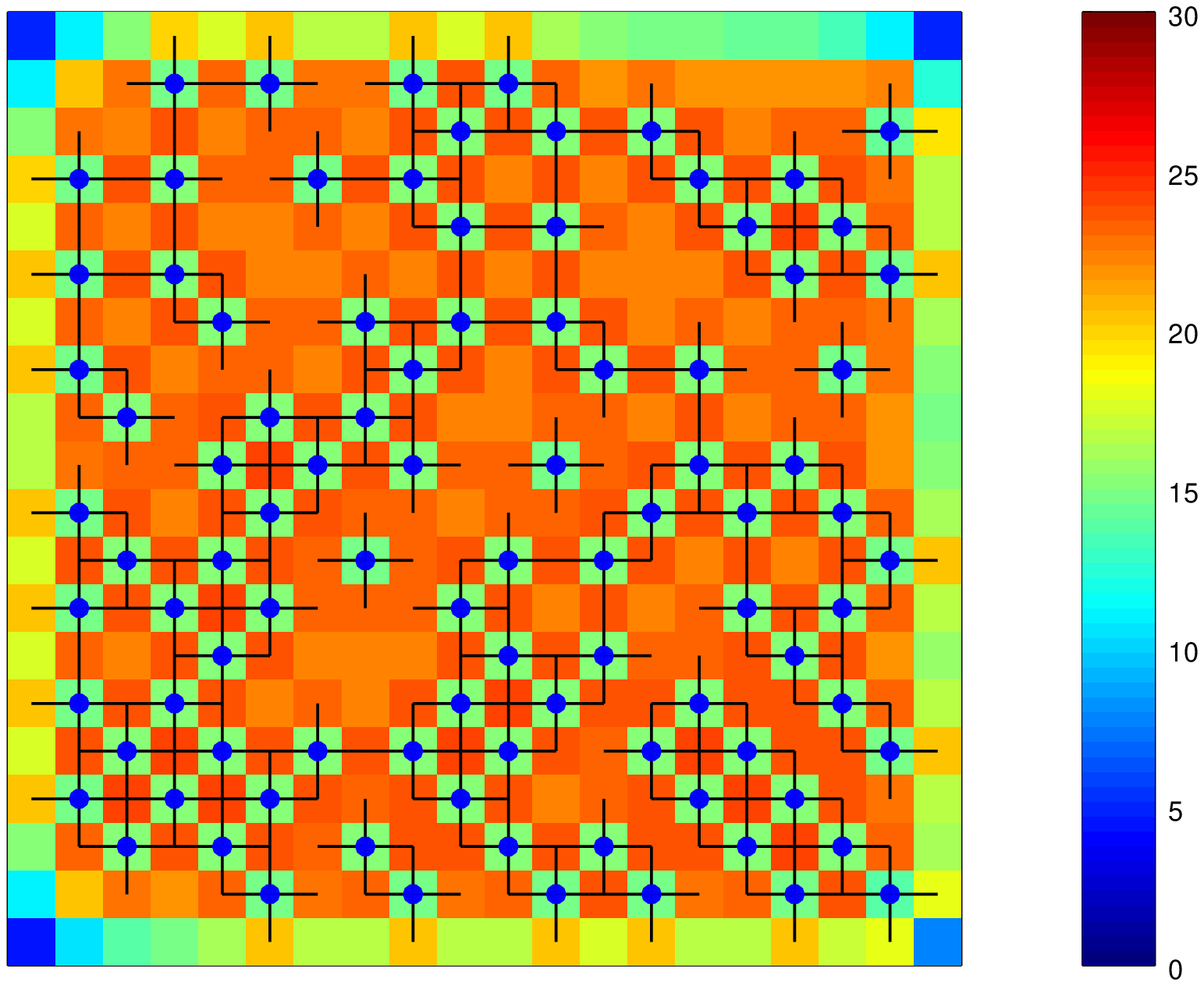}\\
\includegraphics[width=0.9\linewidth]{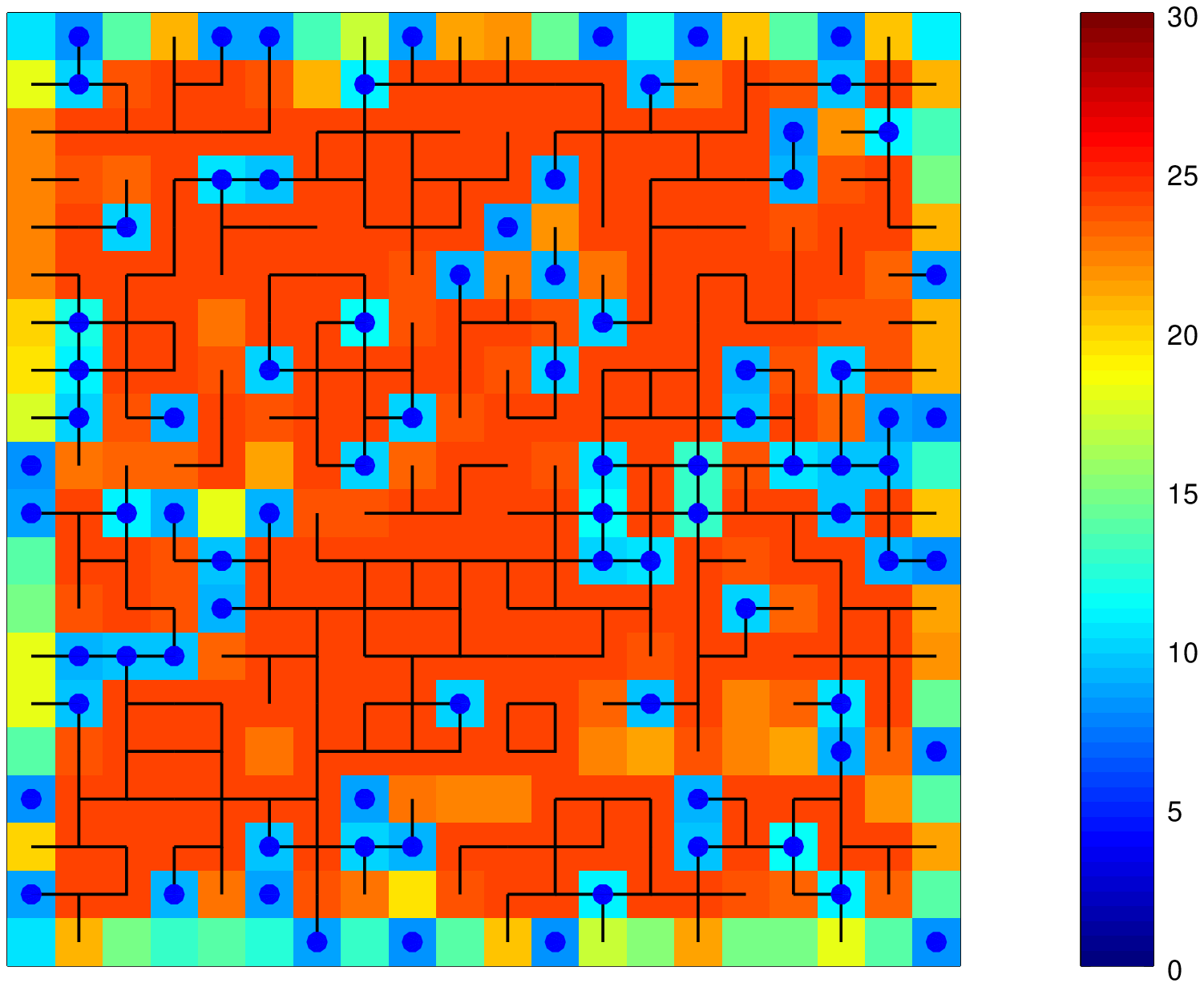}
\caption{\label{fig:denlatt6.6}(Color online) Same as Fig.\
  \ref{fig:denlatt6.6} for $\Delta/t=6.6$.}
\end{figure}

\begin{figure}
\centering
\includegraphics[width=0.9\linewidth]{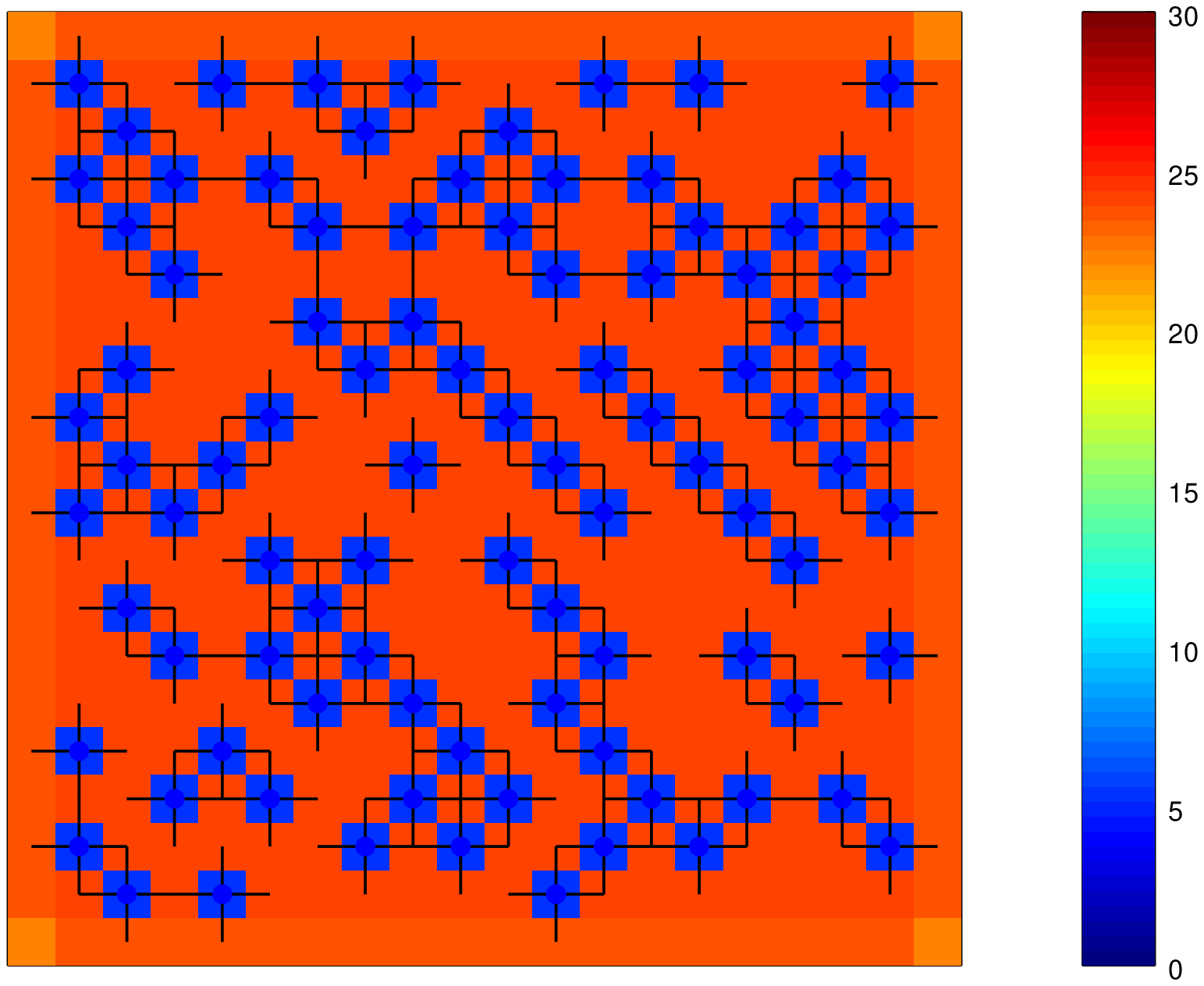}\\
\includegraphics[width=0.9\linewidth]{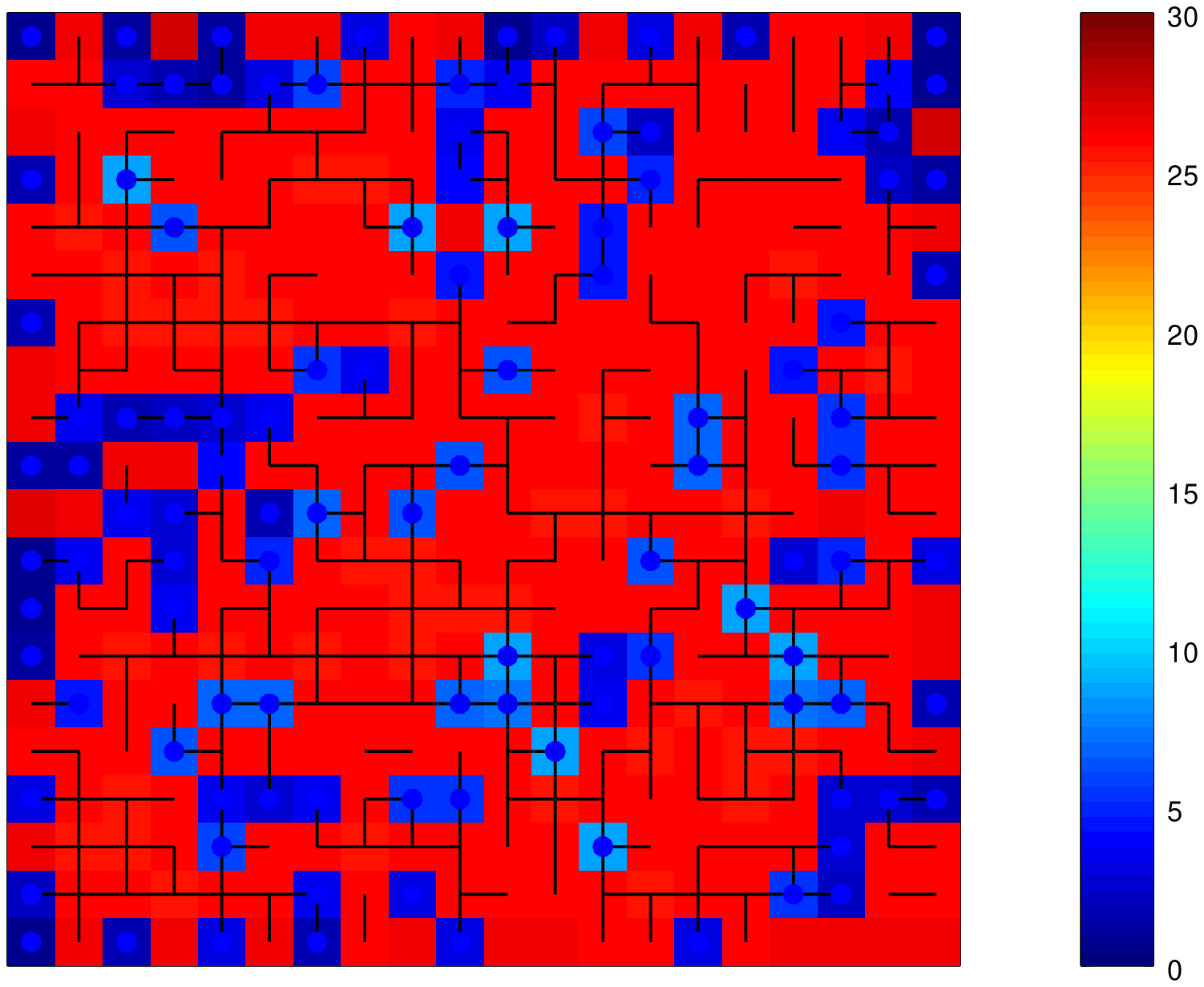}
\caption{\label{fig:denlatt15.1}(Color online) Same as Fig.\
  \ref{fig:denlatt6.6} for $\Delta/t=15.1$.}
\end{figure}

With the aim of characterizing the ground state configurations in the
different regions, we show in Figs.
\ref{fig:denlatt2.1}--\ref{fig:denlatt15.1} the spatial density
distribution $n_i$ for $L=20$, $n=20$ at $\Delta/t\simeq 2$ 
(Fig.~\ref{fig:denlatt2.1}),
$\Delta/t \simeq 6.6$ (Fig.~\ref{fig:denlatt6.6}), 
and $\Delta/t\simeq 15$ (Fig.~\ref{fig:denlatt15.1}) together with 
a pattern showing the
locations of impurities.

The addition of a hopping term $t'$ favors the delocalization of the
density both for the 2D-DRDM and UN-RAND disorders. However, in the
case of the 2D-DRDM, it is more beneficial as it tends to partially
compensate the decrease in the density caused by the site impurity,
reducing the decrease by means of the structured disorder.  For small
values of $\Delta$ (see Fig.\ \ref{fig:denlatt2.1}), in the region
where the effect of $t'$ is dominant, the density in the impurity
regions is even larger with respect to the density elsewhere.  For
large values of $\Delta$ (see Fig.\ \ref{fig:denlatt15.1}), the effect
of both types of disorder is similar as the change in the on-site
energies dominates. This limit gives rise to a strongly depleted
density at the impurity location plus a rather uniform background.
The largest differences among the 2D-DRDM and UN-RAND results are seen
at the minimum of $\mathcal{I}_{\mathrm{GS}}$ (see Fig.\
\ref{fig:denlatt6.6}), where we can clearly observe a more homogeneous
density spread over the lattice (lower $\mathcal{I}_{\mathrm{GS}}$),
and a consequently larger delocalization for the 2D-DRDM than for the
UN-RAND potential.

\begin{figure}
\includegraphics[width=0.9\linewidth]{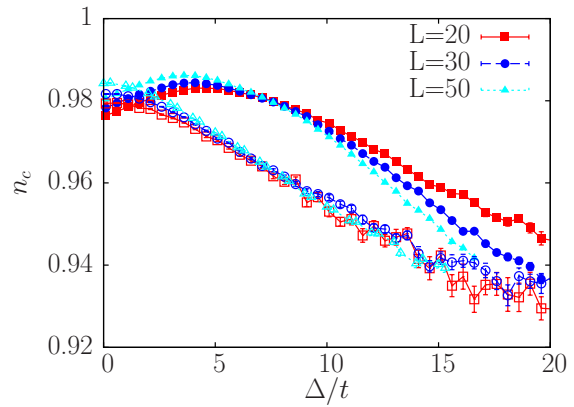}
\caption{\label{fig:N0_L}(Color online) Condensate fraction $n_c$ as a
  function of $\Delta/t$ for $t'/t=2$, $U/t=10^{-2}$ and $n=20$
  particles per site. The different curves correspond to different
  values of $L$ as indicated in the figure.  The filled symbols
  correspond to the 2D-DRDM potential and the empty symbols correspond
  to the UN-RAND potential.}
\end{figure}

The density behavior determines the condensate fraction
$n_c=\sum_i|\langle \Phi_{\mathrm{GS}}|a_i |\Phi_{\mathrm{GS}} \rangle|^2/n$,
as shown in Fig. \ref{fig:N0_L}.
In correspondence of the minimum of the function
$\mathcal{I}_{\mathrm{GS}}$, we observe that the condensate fraction
$n_c$ does not depends of the system size, in the presence of the
2D-DRDM potential. The resonance condition minimizes the
fluctuations with respect the chosen homogeneous basis
$|\psi_i\rangle$ and fixes $n_c$.  
At lower value of $\Delta$, we observe
a ``super-delocalization'' ($\mathcal{I}_{\mathrm{GS}}L^2$ scales as
$L^{-\epsilon}$), and for both the 2D-DRDM and the UN-RAND potentials,
the large value of $t'$ enhances the coherence and $n_c$ increases
with system size. 

At larger values of $\Delta$, where
$\mathcal{I}_{\mathrm{GS}}L^2$ scales as $L^{\epsilon'}$, the 2D-DRDM
impurities create holes in the system, and $n_c$ decreases with system
size.  For the case of the UN-RAND potential, one can observe a
monotonic behavior of $n_c$ as a function of $\Delta$. As for the case
of the 2D-DRDM, the region where all the curves
$\mathcal{I}_{\mathrm{GS}}L^2$ collapse together corresponds to a
region where $n_c$ does not depend on the system size. The difference
with the 2D-DRDM is a larger decrease of $n_c$ in this
region. For 2D-DRDM, only one value of $\Delta$ has this peculiarity,
and the maximum position of the condensate fraction foregoes this
point. Let us remark
that the minimum of $\mathcal{I}_{GS}$ corresponds to the minimum deviation
with respect a homogeneous condensate, and, because of border effects, 
this target state is not necessarily the one that ensures a maximum
value of $n_c$ in finite systems.

The predicted condensate fraction enhancement for the DRDM at low $\Delta$,
being it very small, could be very difficult to be measured.
However the non-diminishing of the coherence in a range
of about $5\Delta$ should be observable, and could be directly
compared with the result for the UN-RAND where the decrease of the
coherence should be sizable.

\section{Conclusions}
\label{sec:concl}
In summary, we introduce a correlated disorder model that is the
natural extension of the DRDM in 2D. We show that, in the
non-interacting regime, such a disorder introduces some delocalized
states if the resonance energy characterizing these structures belongs
to the spectrum of the unperturbed system.  In the presence of weak
interactions, the 2D-DRDM drives the density spatial fluctuations.  By
means of a mode-matching argument that includes the effect of the
interactions, we show that the resonance energy is at the origin of
these phenomena. A direct consequence is a non-monotonic behavior of
the condensate fraction as a function of the disorder strength, and
its enhancement for values close to the resonance condition.  This
work shows that short-range correlations in a disordered potential can
modify and enhance the coherence of a many-body system in the weak
interacting regime. Such effects could be measured in the context of
ultracold atoms with an accurate measurement of the density and
coherence, via, for instance, a fringes contrast interference
experiment. Our results could also be extended to homogeneous systems
provided one is able to engineer suitable impurities that are
transparent for a given energy.

\begin{acknowledgments}
  This work was supported by CNRS PICS grant No. 05922.  PC
  acknowledges partial financial support from ANPCyT grant PICT 2011-01217. 
\end{acknowledgments}

\end{document}